# Effects of GaAs buffer layer on quantum anomalous Hall insulator $V_y(Bi_xSb_{1-x})_{2-y}Te_3$

Yusuke Nakazawa[a)], Takafumi Akiho, Kiyoshi Kanisawa, Hiroshi Irie, Norio Kumada, and Koji Muraki

*NTT Basic Research Laboratories, NTT Corporation, 3-1 Morinosato-Wakamiya, Atsugi 243-0198, Japan*

[a)]Author to whom correspondence should be addressed: yusuke.nakazawa@ntt.com

We report the growth, structural characterization, and transport properties of the quantum anomalous Hall insulator $V_y(Bi_xSb_{1-x})_{2-y}Te_3$ (VBST) grown on a GaAs buffer layer by molecular beam epitaxy on a GaAs(111)A substrate. X-ray diffraction and transmission electron microscopy show that the implementation of a GaAs buffer layer improves the crystal and interface quality compared to the control sample grown directly on an InP substrate. Both samples exhibit the quantum anomalous Hall effect (QAHE), but with similar thermal stability despite the different structural properties. Notably, the QAHE in the sample grown on a GaAs buffer layer displays a significantly larger (almost double) coercive field with a much smaller resistivity peak at the magnetization reversal. Possible effects of the interface quality on the magnetic properties of VBST and the QAHE are discussed.

The quantum anomalous Hall effect (QAHE), characterized by quantized Hall resistance $R_{yx} = \pm h/e^2$ in the absence of external magnetic fields[1,2], represents a class of topologically nontrivial states of matter with spontaneously broken time-reversal symmetry. Among the various material systems that have been found to exhibit the QAHE[3–7], thin films of three-dimensional topological insulators $(Bi_xSb_{1-x})_2Te_3$ doped with Cr or V[3,4] have been the most studied system since the first experimental observation of the QAHE therein[3]. In $Cr_y(Bi_xSb_{1-x})_{2-y}Te_3$ (CBST) and $V_y(Bi_xSb_{1-x})_{2-y}Te_3$ (VBST), when the exchange energy due to spontaneous magnetization exceeds the hybridization energy of the top and bottom surfaces[8], the gap in the surface spectrum becomes nontrivial, where gapless chiral edge states responsible for the QAHE emerge. Extensive research has been conducted on these compounds from a physical point of view, including chiral edge conduction[9–12], universal scaling of the quantum phase transition[13–19], and magnetization mechanisms[20–23], as well as from an application point of view, such as the resistance standard[24–26].

An important issue common to these studies is that the observation of the QAHE requires very low temperatures, typically below ~ 0.1 K[27–29], despite the Curie temperature being on the order of 10 K[3,4]. Various factors, including magnetic disorder[30], inhomogeneity in the Bi/Sb composition[27], and the presence of superparamagnetic domains[31,32], have been considered as sources of bulk dissipation that is triggered by even a slight increase in temperature. To overcome this problem, several approaches using modified layer structure or composition, such as the heterostructure with magnetic modulation doping[27], co-doping of Cr and V[28], and a magnetic capping layer[29], have been investigated and shown to be effective in stabilizing the QAHE up to a few hundred mK or even a few K. On the other hand, previous studies suggest that the robustness of the QAHE is not significantly affected by the choice of substrate. In fact, QAHEs have been observed for CBST and VBST films grown on a variety of substrates, with much the same thermal stability. These include $SrTiO_3(111)$[3,10,12,18,23,28,31,33], $InP(111)A$[11,14,16,17,26,27,34,35], $GaAs(111)B$[9,24,25], $Si(111)$[16,19], C-sapphire[29], and mica[29,36]. We note, however, that these films were grown directly on a substrate. While the use of a buffer layer is known to improve the interface quality and thus the transport properties in undoped topological insulator thin films[37–39], it remains unexplored for magnetically doped films and little is known about its effects on the magnetic properties and the QAHE.

Here, we report the molecular beam epitaxy (MBE) of VBST on a GaAs buffer layer grown on GaAs(111)A substrate and its structural and transport characterization including QAHE. Comparison with VBST grown directly on InP substrate using X-ray diffraction (XRD) and scanning transmission electron microscopy (STEM) shows improved crystal and interface quality due to the implementation of the GaAs buffer layer. Both VBST films exhibit QAHE, but with similar thermal stability despite the difference in crystal/interface quality. However, we found an unexpected behavior in the QAHE of the VBST grown on the GaAs buffer layer; it exhibits a significantly larger coercive field of 1.34 T compared to 0.72 T for the VBST on InP. At the same time, the resistivity peaks at the magnetization reversals are much smaller, suggesting that the dimensionality of the QAHE is different for the two films despite the same thickness. We discuss possible effects of the interface quality on the quantum anomalous Hall state.

We used a multi-chamber MBE system to grow the GaAs buffer layer and VBST in separate growth chambers connected by an ultra-high vacuum transfer line. We used an on-axis GaAs(111)A substrate, for which an atomically flat GaAs layer can be obtained under optimized growth conditions (see the supplementary material for details on GaAs buffer growth)[40–42]. After the buffer layer growth, the sample was transferred to the VBST growth chamber and heated to a growth temperature of $T_g = 150°C$, and then Bi, Sb, V, and Te were co-evaporated. After the VBST growth, the sample was annealed in situ at 250°C for 30 minutes under Te flux. We also grew control samples with VBST directly on InP(111)A substrates. Following previous reports, the InP substrate was preannealed at 380°C for 30 minutes in the VBST growth chamber[40,41] prior to the VBST growth at 150°C - 160°C[43,44]. For all samples, the Bi composition $x = P_{Bi}/(P_{Sb} + P_{Bi})$ was set to 0.35 to tune the chemical potential close to the charge neutrality point[45] and $P_V/(P_{Sb} + P_{Bi} + P_V)$ was set to 0.025 (i.e. $y = 0.050$), where $P_{Sb}$, $P_{Bi}$, and $P_V$ are the beam-equivalent pressures for each element. The beam flux ratio of Te relative to Bi, Sb, and V was set higher than 10 to suppress Te vacancies. Cracker cells were used to evaporate Sb and Te with the cracking temperature set at 600°C and 1,000°C, respectively, while effusion cells were used for Bi and V. We grew VBST films with a design thickness of 7 quintuple layers (QL, 1 QL ≒ 1 nm) at a growth rate of approximately 0.08 QL/min, without a capping layer. Structural properties of the grown films were characterized by XRD and STEM EDX, and transport measurements were performed in a dilution refrigerator with magnetic fields up to 3 T. The Hall and longitudinal resistivities ($\rho_{yx}$ and $\rho_{xx}$) were measured with the samples cut into 1-mm wide pieces and Ohmic contacts made with silver paint.

First, we examine the surfaces of the buffer layer and the VBST. Figure 1(a) shows a reflection high-energy electron diffraction (RHEED) pattern after the buffer layer growth taken along the GaAs($1\bar{1}0$) azimuthal direction. The sharp diffraction spots on the higher-order Laue circles with Kikuchi lines represent good crystallinity and flatness of the buffer layer surface. The atomic force microscopy (AFM) image of the buffer layer surface [Fig. 1(b)] shows a root mean square (RMS) roughness of 0.29 nm, which is a near-optimal value achievable for an MBE-grown GaAs(111)A plane[46]. The step height of 0.3 nm [Fig. 1(c)] corresponds to the lattice spacing of GaAs along the (111) direction. Figure 1(d) is a RHEED pattern for the VBST grown on a GaAs buffer, taken after annealing. The streak pattern confirms a smooth surface. AFM of the VBST surface shows an RMS roughness of 0.44 nm [Figs. 1(e) and 1(f)], which is better than 0.72 nm for the VBST on the InP substrate (see the supplementary material for the AFM of VBST grown on the InP substrate). We note that the VBST layer is much rougher when it is grown directly on a GaAs(111)A substrate without the buffer (see the supplementary material for direct growth on a GaAs(111)A substrate).

We used XRD to compare the crystallinity of the VBST samples grown on the GaAs buffer layer and the InP substrate. For both the samples, diffraction peaks from the (003) series of VBST are clearly visible in the $\theta$-$2\theta$ scans [Fig. 2(a)], confirming their single-crystalline nature. On the other hand, the Laue oscillations around the diffraction peaks are only visible for the sample grown on a GaAs buffer layer, showing its better flatness. The VBST thickness derived from the oscillation period is 6.7 nm, which is consistent with the designed thickness of 7 QL. Furthermore, the VBST(006) diffraction peak in the rocking curve [Fig. 2(b)] has a narrow full width at half maximum of 0.39° for the VBST on the GaAs buffer layer compared to 1.3° on the InP substrate, demonstrating that better crystallinity can be achieved by incorporating a GaAs buffer layer.

Next, we present STEM characterization of the VBST samples to investigate their interface as well as crystal quality at the atomic level. Figures 3(a) and 3(b) are cross-sectional high-angle annular dark field STEM (HAADF-STEM) images of the VBST samples grown on the GaAs buffer layer and the InP substrate, respectively. Notably, QLs are clearly resolved over the entire measurement area for VBST on GaAs buffer layer, while they are not resolved in some areas, indicating poor local crystallinity, for VBST on InP substrate. In addition, there is an inhomogeneous intermediate layer at the interface with the InP substrate. EDX indicates that this inhomogeneous layer is due to V accumulation (not shown). In contrast, the interface with the GaAs buffer is sharp with no transition region. Such unintentional accumulation of magnetic dopants at the substrate interface has been commonly observed in CBST films in previous reports[33,34]. Dopant accumulation may be due to incomplete oxide removal from the substrate, a problem often unavoidable in direct growth on a substrate. This, in turn, suggests that it can be overcome by implementing a buffer layer, as we have shown here. The sharp interface is considered responsible for the Laue oscillations in the XRD $\theta$-$2\theta$ scan observed only on the GaAs buffer. The difference in the interface quality can also be presented as depth profiles of the in-plane lattice constant calculated from the Fourier transform of the STEM images [Figs. 3(c) and 3(d)]. The in-plane lattice constant changes abruptly at the interface with the GaAs buffer, while it changes gradually across the intermediate layer for VBST on InP substrate.

We now present the results of the transport measurements and compare the properties of the QAHEs. Figures 4(a) and 4(b) show the Hall and longitudinal resistivity ($\rho_{yx}$ and $\rho_{xx}$) as a function of the magnetic field at 20 mK. Both samples exhibit the QAHE ($\rho_{yx} = \pm h/e^2$ and $\rho_{xx} = 0$) without gating, indicating that the chemical potential is located near the charge neutrality point. Unexpectedly, we first notice that the coercive field $B_c$ is significantly different. The $B_c$ = 1.34 T for VBST on GaAs buffer is nearly double the $B_c$ = 0.72 T for VBST on InP substrate. Particularly, the $B_c$ = 1.34 T for the VBST on GaAs buffer is the largest among those reported for VBST. This coercive field enhancement is observed in all VBST samples we grew on the GaAs buffer layer (not shown). In addition to $B_c$, we notice that the height of the $\rho_{xx}$ peaks at the magnetization reversal is also significantly different, with VBST on GaAs buffer showing a much smaller $\rho_{xx}$ peak. Possible impacts of the GaAs buffer layer on the coercive field and the $\rho_{xx}$ peak value are discussed later. In contrast to the significantly different behavior of the magnetization reversal at $B_c$, the behavior around $B = 0$ T is surprisingly similar between the two samples. Crossing $B = 0$ T causes $\rho_{yx}$ and $\rho_{xx}$ to deviate from the plateau values over a finite range of $B$, a behavior observed in previous reports and considered to be a manifestation of the superparamagnetism of magnetic domains[32].

Figures 4(c) and 4(d) present the temperature dependence of $\rho_{yx}$ and $\rho_{xx}$ at $B = 0$ T, measured after applying $B = 3$ T to saturate the magnetization. The two samples are comparable in the robustness of the QAHE, with the quantization persisting only up to $T = 0.1$ K, similar to previous reports on uniformly doped VBST films[43]. This result indicates that the robustness of the QAHE is not constrained by crystallinity or interface quality (see the supplementary material for the complete datasets).

Finally, we discuss the origin of the different coercive fields of VBST films grown on a GaAs buffer layer and directly on an InP substrate. We note that previous reports have observed superparamagnetic behavior of magnetic domains upon magnetization reversal in CBST films[46]. According to the Stoner-Wohlfarth model[47], the coercive field of superparamagnetic domains is proportional to $K/M_s$, where $K$ is the magnetic anisotropy energy and $M_s$ is the saturation magnetization per unit volume. Therefore, a higher $K$ and/or lower $M_s$ will result in a high $B_c$. Magnetic anisotropy is contributed by two main sources, magnetocrystalline anisotropy and surface/interface anisotropy, the latter becoming important in thin films. The difference in the crystallinity, as manifested in the X-ray rocking curve and STEM, respectively, can affect magnetocrystalline anisotropy. Different strain distributions, as inferred from the in-plane lattice constant profiles, may be another cause. Additionally, in samples grown directly on InP substrates, V accumulation at the substrate interface would affect $K$ in two ways. First, it would make the actual V composition in the bulk of the VBST layer less than that expected from the V flux ratio. Since a previous study on CBST reports that the coercive field increases with $y$[14], it is possible that the reduced V composition decreases $B_c$. Second, the presence of the intermediate layer would alter the substrate interface contribution to $K$.

The above speculation of altered V composition also explains the different $\rho_{xx}$ peak heights. A previous study shows that thinner (thicker) VBST samples exhibit higher (lower) $\rho_{xx}$ peaks at the magnetization reversal, following the semicircle law that reflects the dimensionality of the film[19]. Our result that the samples with nominally the same thickness show different $\rho_{xx}$ peak heights requires another explanation. We note that the dimensionality of a magnetic topological insulator film is determined by the interplay between the hybridization energy of the top and bottom surfaces and the exchange energy, with the film thickness tuning the former. The reduced V composition would make the exchange energy smaller and the VBST film effectively thinner. Thus, the behavior of the $\rho_{xx}$ peak reinforces the scenario of an altered V composition in the bulk of the VBST layer.

In summary, we studied the effects of a GaAs buffer layer on the structural and transport properties of the quantum anomalous Hall insulator VBST. XRD and STEM EDX analysis demonstrated improved crystal and interface quality compared to direct growth on an InP substrate. While the robustness of the QAHE against temperature remained almost unchanged, we found the coercive field to be significantly enhanced, indicating that the GaAs buffer layer significantly affects the magnetic property of VBST. The obtained $B_c$ of 1.34 T is the highest among the VBST films reported to exhibit the QAHE. This provides an insight into the relation between the structural and magnetic properties of magnetic topological insulators, and demonstrates that a buffer layer is an additional tool to control them, which will be useful for engineering quantum anomalous Hall insulators and their heterostructures with controlled structural and magnetic properties.

See the supplementary material for additional information on the GaAs buffer and VBST growth, VBST direct growth on a GaAs(111)A substrate, AFM and STEM of the VBST grown on the InP(111)A substrate, and the temperature dependence of the QAHEs.

## AUTHOR DECLARATIONS

### Conflict of Interest

The authors have no conflicts to disclose.

### Author Contributions

**Yusuke Nakazawa:** Conceptualization (equal); data curation (lead); formal analysis (lead); methodology (equal); visualization (lead); writing - original draft (lead); writing - review and editing (equal). **Takafumi Akiho:** Conceptualization (equal); methodology (equal); writing - review and editing (equal). **Kiyoshi Kanisawa:** Methodology (equal); writing - review and editing (equal). **Hiroshi Irie:** Methodology (equal); writing - review and editing (equal). **Norio Kumada:** Supervision (equal); writing - review and editing (equal). **Koji Muraki:** Conceptualization (equal); formal analysis; (supporting); methodology (equal); supervision (equal); writing - review and editing (equal).

## DATA AVAILABILITY

The data that support the findings of this study are available from the corresponding author upon reasonable request.

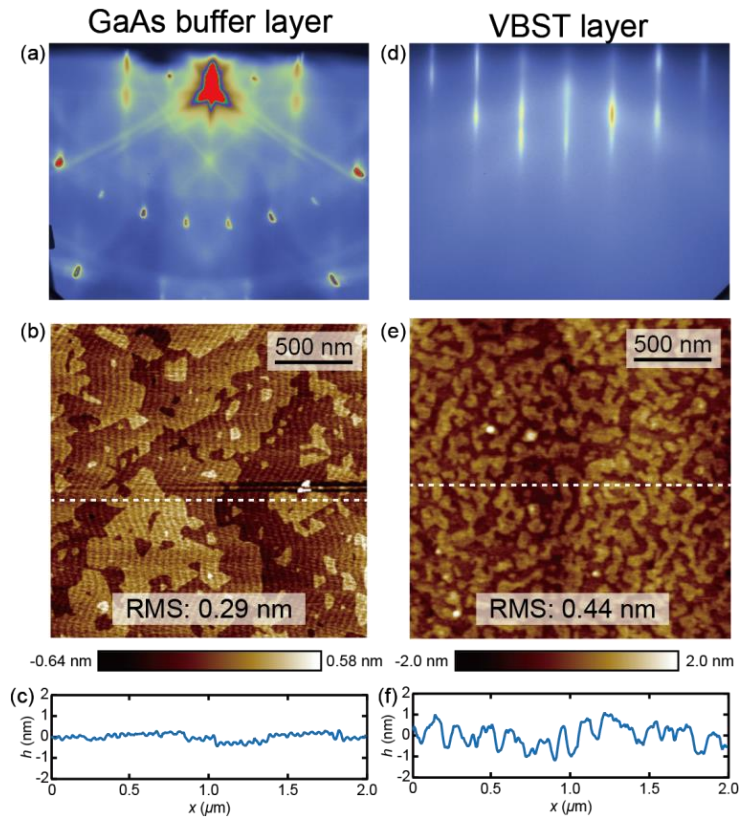

FIG. 1. (a) RHEED pattern of a GaAs buffer layer taken along the ($1\bar{1}0$) azimuthal direction. (b) AFM image and (c) height profile of the GaAs buffer layer. The height profile is measured along the dashed line in (b). Periodic wavy pattern in the AFM image is instrumental noise. (d) RHEED pattern of a VBST layer grown on the GaAs buffer layer taken along the VBST(210) azimuth. (e) AFM image and (f) height profile of the VBST layer.

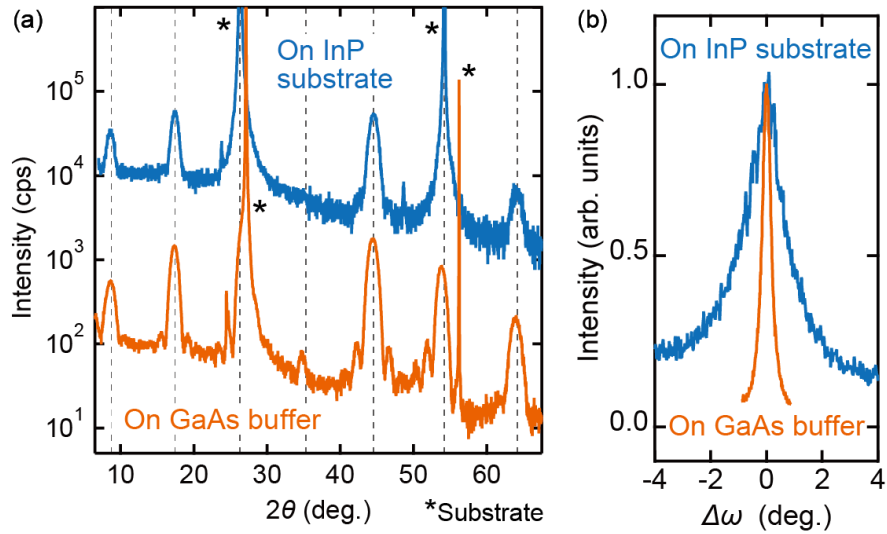

FIG. 2. (a) XRD $\theta$-$2\theta$ scans of the VBST films grown on the InP substrate ($T_g$ = 150°C) and the GaAs buffer layer. The dashed lines show diffraction peak positions for the VBST(003) series. The asterisks denote diffraction peaks from the substrates. The plots are offset for clarity. (b) Rocking curves for the VBST (006) diffraction.

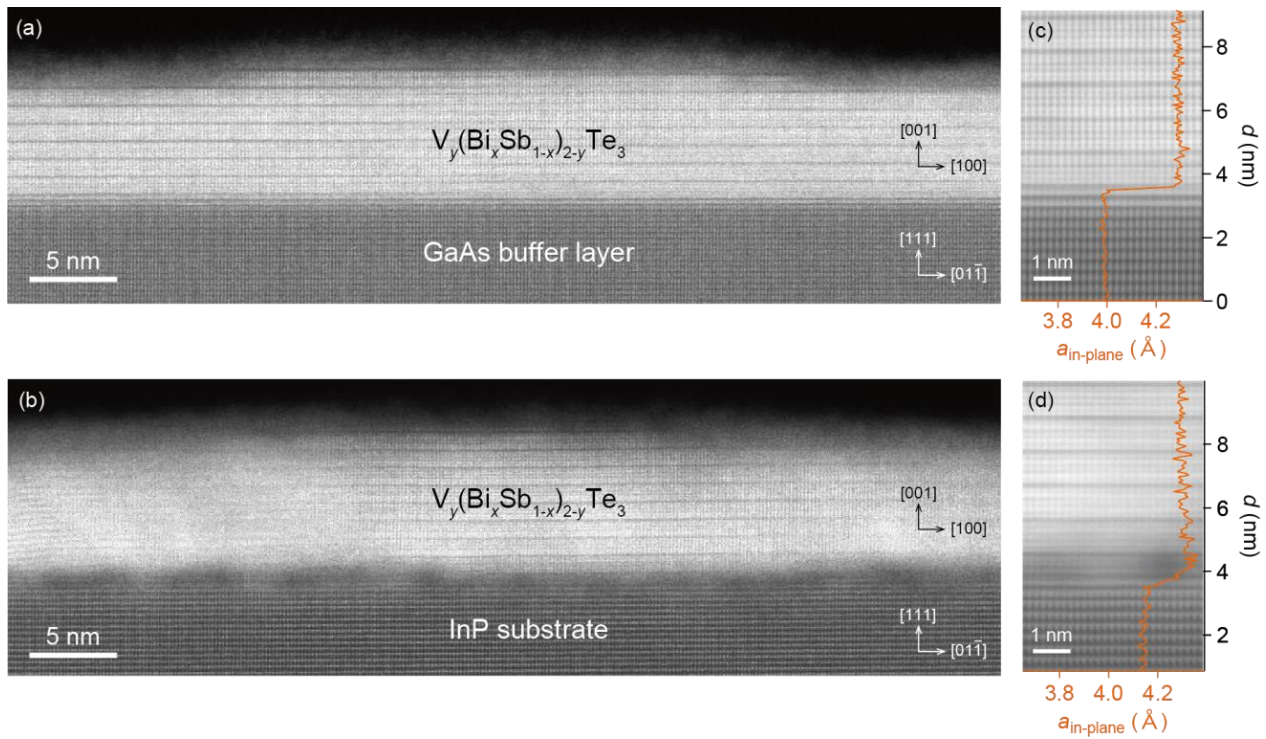

FIG. 3. (a), (b) Cross-sectional HAADF-STEM images for the VBST films (a) grown on the GaAs buffer layer and (b) directly on the InP substrate ($T_g$ = 150°C), respectively. (c), (d) Depth profiles of in-plane lattice constants $a_{\text{in-plane}}$ calculated from the Fourier transformation of the STEM images.

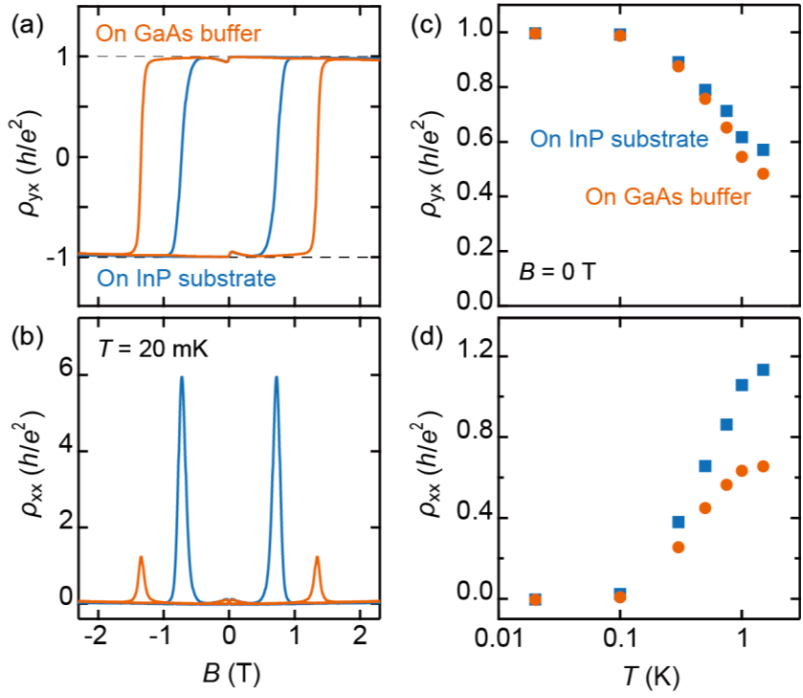

FIG. 4. (a) Magnetic field dependence of the Hall ($\rho_{yx}$) and (b) longitudinal ($\rho_{xx}$) resistivity for the VBST samples grown on the GaAs buffer layer and the InP substrate ($T_g = 160°C$) at $T = 20$ mK. (c), (d) Temperature dependence of (c) $\rho_{yx}$ and (b) $\rho_{xx}$ at $B = 0$ T after applying $B = 3$ T to saturate the magnetization.

# Supplementary materials for

Effects of GaAs buffer layer on quantum anomalous Hall insulator $V_y(Bi_xSb_{1-x})_{2-y}Te_3$

Yusuke Nakazawa[a)], Takafumi Akiho, Kiyoshi Kanisawa, Hiroshi Irie,
Norio Kumada, and Koji Muraki

*NTT Basic Research Laboratories, NTT Corporation, 3-1 Morinosato-Wakamiya, Atsugi 243-0198, Japan*

[a)]Author to whom correspondence should be addressed: yusuke.nakazawa@ntt.com

## S1. Growth of the GaAs buffer and the VBST layers

The GaAs buffer layer and the subsequent VBST were grown with separate Sienta Omicron EVO50 MBE modules connected with an UHV transfer line. The GaAs(111)A substrate was attached to a 2-inch sapphire wafer with In-Ga solder, and they were mounted over a hole in a Mo block, which enabled efficient substrate heating with a radiation heater. Loaded in a growth chamber, the substrate was annealed under $As_4$ flux to desorb native oxide. The oxide desorption was confirmed as a change in the RHEED pattern, and the GaAs growth temperature was set 50°C below the oxide desorption point. For the buffer layer growth, the V/III ratio determined by the incorporation rates of Ga and As was set to 11.4. Thickness and the growth rate of the GaAs buffer layer are 200 nm and 0.28 Å/s, respectively. After the GaAs buffer growth, as described in the main text, the sample was transferred to the VBST growth chamber under UHV, and the VBST layer was grown by co-evaporation of Bi (6N, Osaka Asahi), Sb (7N, DOWA Electronics), V (4N, American Elements), and Te (7N, Osaka Asahi) flux. For the cracking temperatures of Sb and Te, the beam flux is considered to be dominated by $Sb_4$ tetramers and atomic Te, respectively[1,2].

## S2. VBST growth on a GaAs(111)A substrate without the buffer layer

We also prepared a VBST sample grown directly on a GaAs(111)A substrate without the buffer layer, for comparison. After the native oxide desorption, the GaAs(111)A substrate was transferred to the VBST chamber, and then VBST was grown on it under the same growth condition as that grown with the buffer layer. Figure S1(a) is an RHEED pattern during the VBST growth directly on the GaAs substrate.

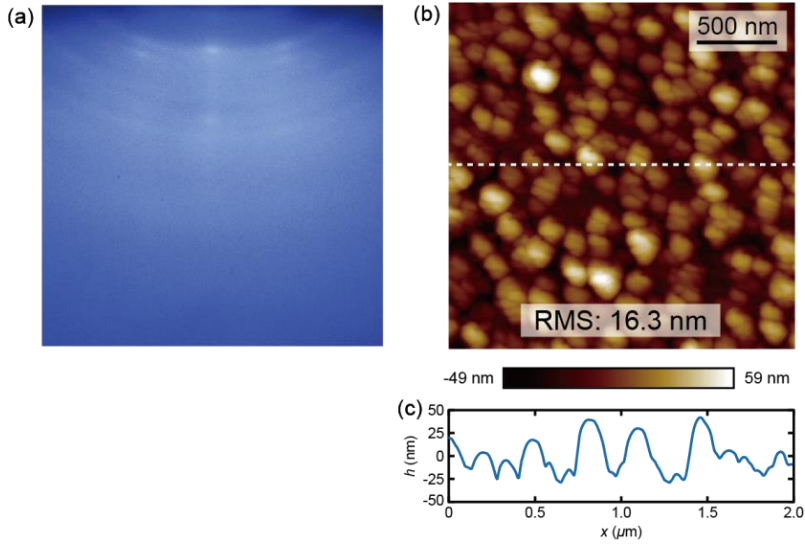

FIG. S1. (a) RHEED and (b) AFM images of a VBST sample directly grown on a GaAs(111)A substrate without a buffer layer, and (c) AFM height profile of the VBST sample.

In contrast to the streak pattern observed for the VBST grown on the buffer layer [Fig. 1(d) in the main text], a ring pattern was observed, indicating rough VBST surface. Rough surface morphology was confirmed by the AFM as shown in Figs. S1(b) and S1(c). The RMS roughness of 16.3 nm is much larger than that grown with the buffer layer. This VBST sample is highly resistive, and the transport properties could not be measured since the film is not continuous because of the three-dimensional growth mode.

## S3. AFM image of the VBST sample grown on the InP(111)A substrate

Figures S2(a) and S2(b) show an AFM image of the VBST sample grown on the InP substrate and its height profile. Compared to the AFM image for the VBST grown on the GaAs buffer layer [Fig. 1(e) in the main text], the terrace structure is not clearly resolved, and the RMS roughness is larger. In addition, more particles are observed on the VBST surface grown on the InP substrate.

## S4. STEM image for the VBST grown on the InP(111)A substrate at $T_g = 160°C$

Figure S3 is a STEM image for the VBST sample grown on the InP substrate at $T_g = 160°C$. Similar to that for the VBST grown on the InP substrate at $T_g = 150°C$ [Fig. 3(b) in the main text], V accumulation layer is formed at the interface. In addition, the VBST layer thickness is 7 or 8 QLs excluding the V-accumulation layer, which is also similar to the sample grown at $T_g = 150°C$, suggesting that the VBST growth rate does not considerably depend on the growth temperature in this temperature range.

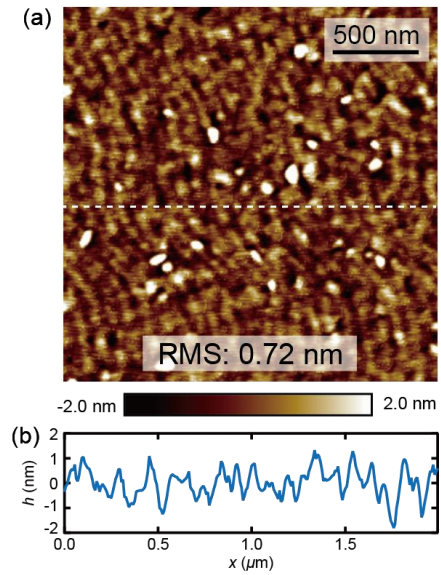

FIG. S2. (a) AFM image of the VBST sample grown on the InP(111)A substrate ($T_g = 150°C$) and (b) its height profile taken along the dotted line in (a). A wavy pattern in the image is instrumental noise.

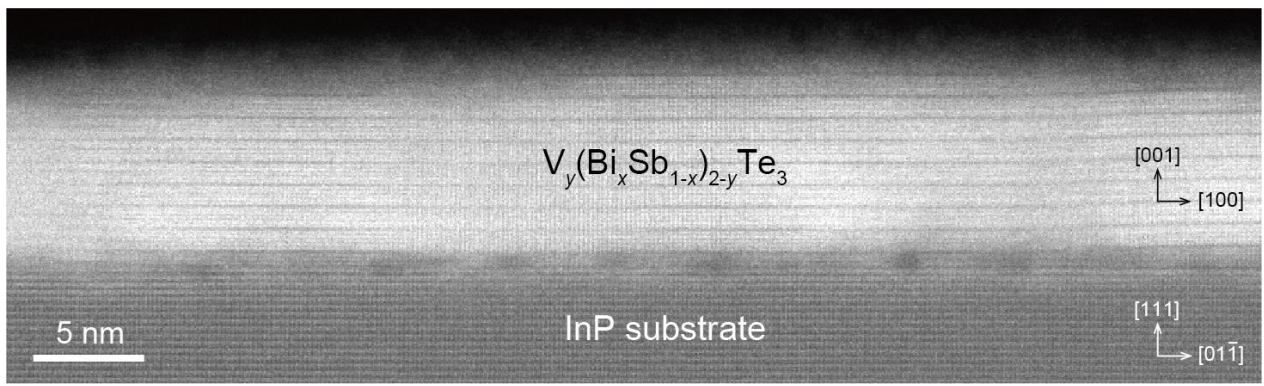

FIG. S3. Cross-sectional HAADF-STEM image for the VBST film grown on the InP substrate at $T_g = 160°C$.

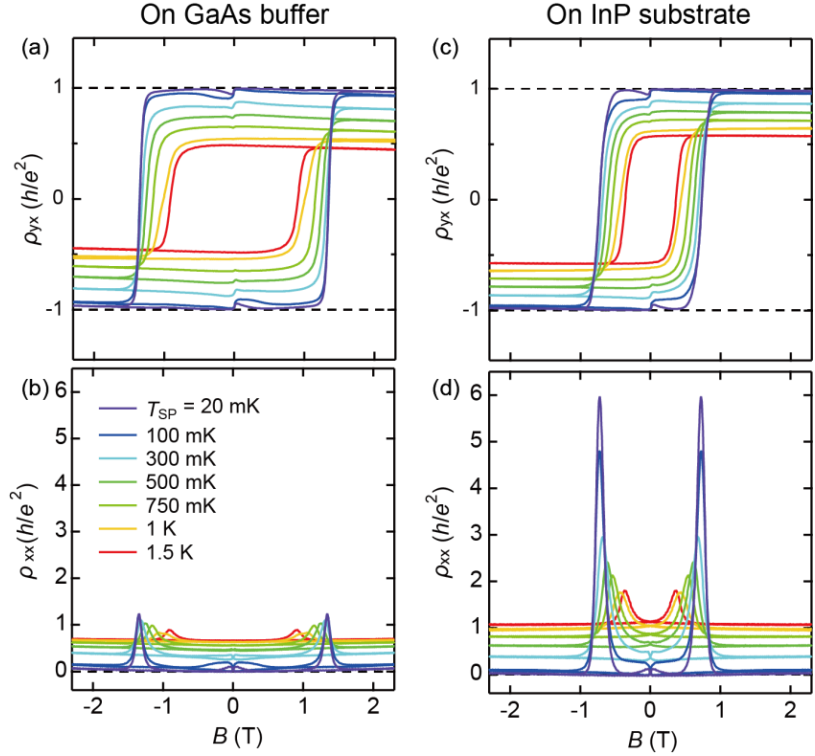

FIG. S4. Magnetic field dependence of $\rho_{yx}$ and $\rho_{xx}$ measured at different temperatures for the VBST samples grown on (a), (b) the GaAs buffer layer and (c), (d) the InP substrate.

## S5. Complete datasets for the temperature dependence of the QAHE

Figure S4 presents complete datasets for the temperature dependence of the QAHE shown in Figs. 4(c) and 4(d) in the main text. The deviations from the plateau values observed when crossing $B = 0$ T are less pronounced in raising the temperature and are not observed for $T \geq 1$ K.

## Supplementary references